# Volume Effects on the Glass Transition Dynamics

C. M. Roland[1,3], K.J. McGrath[1,2], and R. Casalini[1,2,4]

[1]*Naval Research Laboratory, Code 6120, Washington, DC 20375-5342*

[2]*George Mason University, Chemistry Department, Fairfax, VA 22030*

[3] mike.roland@nrl.navy.mil          [4] casalini@ccf.nrl.navy.mil

*(October 26, 2005)*

**Abstract**

The role of jamming (steric constraints) and its relationship to the available volume is addressed by examining the effect that certain modifications of a glass-former have on the ratio of its isochoric and isobaric activation enthalpies. This ratio reflects the relative contribution of volume (density) and temperature (thermal energy) to the temperature-dependence of the relaxation times of liquids and polymers. We find that an increase in the available volume confers a stronger volume-dependence to the relaxation dynamics, a result at odds with free volume interpretations of the glass transition.

## 1. Introduction

Among the many intriguing phenomena of our physical world, the spectacular change in the behavior of liquids undergoing vitrification continues to fascinate observers and inspire researchers. Changes in temperature of only a few degrees can alter the viscosity (or relaxation time) of a material by more than a factor of 1000. The fact that such spectacular changes are accompanied by almost negligible changes in molecular structure or configuration lead naturally to the concept of unoccupied volume as playing a governing role in the dynamics of molecules. However, free volume models *per se* [1,2,3,4,5] have largely fallen out of favor, with recent attempts to interpret the glass transition focused on entropy and the energy landscape as the important aspects [6,7,8,9,10,11,12]. From relaxation measurements at elevated pressure in combination with the equation of state for a glass-former, the relative contribution of volume and temperature can be quantified experimentally from the ratio of the respective activation enthalpies at constant volume and constant pressure [13,14]

$$E_V / E_P = \frac{d \log \tau}{dT^{-1}}\bigg|_V \bigg/ \frac{d \log \tau}{dT^{-1}}\bigg|_P \qquad (1)$$

This ratio, equal to the ratio of the corresponding isochoric and isobaric fragilities, is usually evaluated at the glass transition temperature, $T_g$, at ambient pressure. Results for molecular liquids can be summarized as [15,16]

$$0.38 \leq E_V / E_P \leq 0.64$$

indicating that volume and temperature exert a comparable influence ($E_V / E_P \sim 0.5$). For polymers, the end-to-end distance of the chains is insensitive to pressure [17], since intramolecular bonds are less sensitive to pressure than intermolecular bonds. This makes volume effects weaker, as reflected in larger values of the enthalpy ratio [15,16]

$$0.52 \leq E_V / E_P \leq 0.81$$

(An exceptional polymer is polyphenylene oxide, which, due to its flexible chain structure in combination with a high $T_g$, has an unusually small $E_V/E_P$ =0.25 [18].) These values are for $T \sim T_g$; that is, for relaxation times in the range from 1 to 100 s.

In this work we examine the effect of two changes on the dynamics of glass-formers: the temperature and (for a polymer) the molecular weight. Increases in the former and decreases of the latter both confer more "unoccupied" or free volume and also enhance molecular mobility. However, as seen herein, additional free volume serves to decrease $E_V / E_P$ in both cases, suggesting that conventional free volume ideas are untenable.

## 2. Experimental

Pressure-volume-temperature measurements used a Gnomix apparatus, based on the confining fluid technique (Zoller P and Walsh DJ 1995 *Standard Pressure-Volume-Temperature Data for Polymers* Technomic, Lancaster, PA). Samples were molded into cylinders, then immersed in mercury inside a flexible bellows. Volume changes of the sample were deduced by subtracting the contribution from the mercury. The absolute density was determined at ambient conditions using the buoyancy method.

## 3. Results and Discussion

### 3.1 Temperature and pressure variation of the volume contribution

Master curves of the structural relaxation times measured by dielectric, neutron, or light scattering at various T and P can be obtained by expressing τ as a function of the product of temperature times the specific volume with the latter raised to a material-constant, γ; i.e., $\tau = \Im(TV^{\gamma})$ [19,20]. From this scaling of the glass transition dynamics the follow relation is obtained

$$\left.\frac{E_V}{E_P}\right|_T = \left(1 + \gamma T \alpha_p(T)\right)^{-1} \qquad (2)$$

in which $\alpha_P$ is the isobaric thermal expansion coefficient. Eq.(2) is verified in Figure 1 with data on 18 glass-formers, both molecular and polymeric with 0.13 ≤γ≤0.85 at $T_g$, thus serving to corroborate the empirical $\tau = \Im(T\upsilon^{\gamma})$ scaling.

Since γ is a constant, the variation in the enthalpy ratio can be calculated from the equation of state. Thus, the relative contribution of V and T to the dynamics can be determined for any condition above $T_g$. Direct experimental measurements of the dependence of $E_V/E_P$ on temperature or pressure are rare. Floudas [21] found that the ratio decreased with increasing T for poly(2-vinylpyridine). From eq.(2) and the fact that $\alpha_P$ increases with temperature, we can infer that $E_V/E_P$ will be a decreasing function of temperature. On the other hand, the product $T_g\alpha_P$ gets smaller with increasing pressure, so that $E_V/E_P$ increases with P. Thus, increases in volume, due either to higher T or lower P, increase the influence that volume has on the dynamics.

We show this in Figure 2 with four examples, propylene carbonate (PC), o-terphenyl (OTP), salol (phenyl salicylate), and 1,1'-di(4-methoxy-5-methylphenyl)cyclohexane (BMMPC). The ratio $E_V/E_P$ decreases monotonically with temperature, indicating a monotonic increase in the effect of volume, relative to that of temperature. A diminution in the role of activated transport is expected at higher temperatures, as thermal energies become comparable to and larger than the potential barriers. It is also interesting to note that for three of the liquids in Fig. 2, the data extends beyond the temperature of the dynamic crossover (i.e., $T > T_B$) [22,23,24]. This means that the changes in dynamics at $T_B$ (as reflected in changes in the temperature variation of the relaxation time and the dielectric strength, in the onset of translational-rotational decoupling, in the splitting of the α- and β-processes [25], etc.) are not due to any modification in the qualitative nature of the dynamics. Thus, there is no evidence of percolation of vacancies [2,26] or an onset of landscape-dominated dynamics [27,28,29] at the dynamic crossover.

### 3.2 Effect of Chain Ends (Molecular Weight)

The variation of $E_V/E_P$ among different glass-formers makes clear the influence of chemical structure on the nature of the local dynamics. However, it is of interest to alter the dynamics of a material without changing its chemical structure. The most obvious way to do this is to change the thermodynamic conditions, for example as described in the prior section. Another approach is to measure the segmental relaxation behavior of a polymer as a function of its chain length. Chemically-identical materials differing only in molecular weight, M, will have the same energy barrier to conformational transitions and, apart from the chain ends, the same intermolecular potential. However, it is well known that below some characteristic M, the glass transition temperature begins to decrease [30,31]. Similarly, it has been found that for certain polymers such as polystyrene (PS) [32,33], polypropyleneglycol [34], and polyisobutylene [35], the fragility decreases with decreasing molecular weight.. Herein we determine whether $E_V/E_P$ exhibits a similar dependence on M.

In Figure 3 are shown volume versus temperature data for a PS (weight average M = 13.7 kg/mol) measured at various pressures. The enthalpy ratio can be calculated from the ratio of the isochronal, $\alpha_\tau$ (= $-V^{-1}$ $(\partial V/\partial T)_\tau$), and isobaric, $\alpha_P$ (= $-V^{-1}$ $(\partial V/\partial T)_P$), thermal expansion coefficients using the formula [36]

$$\frac{E_V}{E_P}\bigg|_T = \left(1 - \alpha_P / \alpha_\tau\right)^{-1} \qquad (3)$$

We evaluate eq.(3) at $T_g$, taking advantage of the fact that $\alpha_\tau = \alpha(T_g)$. This follows from the empirical fact that at the $T_g$ determined by PVT measurements $\tau$ is constant, independent of P or V [15]. This is demonstrated in Figure 4 showing data from various sources [37,38,39,40], in which the temperature for a fixed $\tau$ is the same (for all pressures) as the $T_g$ determined by volumetric experiments. The particular value of $\tau$ in Fig. 4 varies in the range from 10 to 40 s, dependent on the rate of temperature change in the volume measurements, as well as the definition of $\tau$. Commonly the latter is taken as the inverse of the (angular) frequency of the peak in the dispersion in the dielectric loss.

The isobaric expansivity at atmospheric pressure is obtained directly from the experimental measurements (as indicated by the solid line in Fig. 3). $T_g$ is determined from the deviation of the experimental volumes from the linear extrapolation of the liquid data. This method minimizes the influence of the fictive temperature of the glass, which depends on thermal history. The $T_g$'s are denoted by filled circles in the figure. We obtain $E_V/E_P = 0.517 \pm 0.015$ for M = 13.7 kg/mol.

Results for the other PS samples are displayed in Figure 5, which also includes a datum from the literature [32] for M = 3.5 kg/mol. There is a systematic increase in the ratio with increasing molecular weight. Thus, a higher concentration of chain ends coincides with a stronger influence of volume of the local on the segmental dynamics.

## 4. Discussion

Historically, there has been a dichotomy in interpretations of the glass transition. The polymer community has embraced primarily free volume models of polymer dynamics [1]. One obvious problem with the free volume concept is that the volume at the glass transition is not a constant (even for a given material – see Fig. 4), contrary to expectations. Another inconsistency is that attempts to use models to quantify the free volume leads to values at odds with estimates from actual PVT measurements [26].

More recently, positron annihilation lifetime spectroscopy (PALS) has been employed to characterize the free volume (unoccupied holes) in glass-forming liquids. Ngai et al. [41] reported that at the dynamic crossover, $T_B$, there is a discontinuity in the magnitude of $\tau_3$, the mean lifetime of the positronium ion, for propylene carbonate, o-terphenyl, glycerol, and propylene glycol. Since $\tau_3$ is a measure of the average free volume hole size, the implication is that the dynamic crossover reflects a qualitative change in the free volume and its distribution. Somewhat similarly, Bartos et al. found that the slope of $\tau_3$ versus T curves for OTP [42] and glycerol [43] systematically change at $T_B$. These PALS results call to mind the free volume model of Grest and Cohen [2], which identifies a characteristic temperature at which the free volume percolates (forms continuous pathways). When the Cohen-Grest expression for $\tau_\alpha$ is fit to experimental data, the characteristic temperature of the model is found to be equal to $T_B$ [26]. The PALS results have also been interpreted in terms of the free volume model of Bendler, Fontanella, and Shlesinger [44], with changes in the $\tau_3(T)$ behavior related to clustering of free volume "defects". However, the PALS data and the various interpretations are at odds with the data in Fig. 2. The relative contribution of volume increases monotonically with temperature without any discontinuity. Thus, the putative percolation and clustering of free volume would have to occur without changing the effect volume has on the relaxation, which is contrary to the underlying idea that free volume governs the dynamics.

Historically, the alternative to a free volume approach is to interpret the dynamics in terms of activated transport or hopping over potential barriers on a free energy landscape [27,45]. Since relaxation times are invariably non-Arrhenius, thermal activation models describe experimental $\tau(T)$ data by invoking an explicit density-dependent activation energy [46,47,48]. The latter immediately confers a dependence on the density, reflected in our scaling law, $\tau = \Im(TV^\gamma)$ [19,20]. It should be emphasized that within the scope of thermally-activated dynamics, volume can still be the dominant control variable rather than temperature (i.e., $E_V/E_P < 0.5$), notwithstanding arguments to the contrary [49,50]. From this perspective, the decrease of $E_V/E_P$ with temperature seen in Fig. 2, reflecting a diminution of the role of temperature on $\tau(T)$, implies that the available thermal energy is becoming substantial relative to the size of the potential energy barriers.

The role of volume is manifest in the strong effect of chain ends on the $\tau(T)$. Chain ends are associated with a greater degree of unoccupied volume; that is, the end units have greater mobility, a packing effect due to their somewhat different chemical structure in comparison to the main repeat units (although large chemical dissimilarities can have the opposite effect [51]). This excess configurational freedom of the chain ends give rise to a molecular weight dependence of $T_g$, described by an equation derived from free volume theory [31]

$$T_g(M)^{-1} = T_g(M = \infty)^{-1} + B/M \qquad (4)$$

where B is a species-specific constant. For PS B = 0.78 and the limiting, high molecular weight value is $T_g$ = 374K [33]. The fragility of PS shows a similar variation with M [32,33].

As seen in Fig. 5, the chain ends and their associated mobility increase the volume dependence of $\tau(T)$. As more free volume becomes available, the dynamics are governed to a greater extent by volume. Like the results in Fig. 2, this is clearly contrary to free volume models. A similar situation has been reported for 1,2-polybutadiene networks [52,53]. Prepared by free-radical crosslinking, the network junctions have a high degree of functionality; that is, a confluence of many chains at the junction. This implies less free volume, as reflected in a systematic increase in $T_g$ of as much as 25 deg with extent of crosslinking [53]. The relative contribution of volume to the dynamics, as seen in the decrease in $E_V/E_P$, decreases with crosslinking.

## 5. Summary

From the scaling relationship $\tau = \Im(TV^\gamma)$, which has been experimentally verified for over 50 polymeric and molecular glass-formers, the relative contribution of volume and temperature to the dynamics is determined for four materials over a range of temperatures in the equilibrium liquid state. In all cases $E_V/E_P$ is found to be a decreasing function of temperature, even for temperature encompassing the dynamic crossover regime. This indicates that as more volume becomes available, the dynamics become more strongly governed by the volume. This is at odds with free volume ideas. The smooth variation of $E_V/E_P$ through the dynamic crossover, at which PALS measurements indicate a change in mean unoccupied hole size, also suggests that free volume has no direct connection to molecular motions. These results are supported by measurements indicating that $E_V/E_P$ also becomes smaller as the molecular weight of PS is reduced; that is, the excess volume conferred by the chain ends magnifies the influence of volume.

**Acknowledgment.**

This work was supported by the Office of Naval Research.

**Figure Captions**

Figure 1. Relaxation parameters plotted in the form of eq.(2) [15]

Figure 2. Temperature dependence of the activation enthalpy, the latter calculated from eq. (2). The glass transition temperature, dynamic crossover temperature, and the temperature at which τ(T) becomes Arrhenius are indicated by arrows. Data are from [15,22,54].

Figure 3. Variation of the specific volume of PS with temperature for P = 10 MPa to (top to bottom). The fits to the liquid data are indicated by the solid line.

Figure 4. Comparison of the glass transition temperature determined from the change in the volume expansivity (solid symbols) to the temperature at which: (○) the dielectric relaxation time equals 10 s for cresolphthalein-dimethylether [37]; (□) the relaxation time measured by dynamic light scattering equals 40 s for diglycidylether of bisphenol A [38]; (△) the dielectric relaxation time equals 100 s for poly(methylmethacrylate) (extrapolated from the data of Theobald et al. [40] (▽) the dielectric relaxation time equals 100 s for p-phenylene [39].

Figure 5. Variation of activation enthalpy ratio with polystyrene molecular weight. Typical uncertainty is indicated by the error bar.

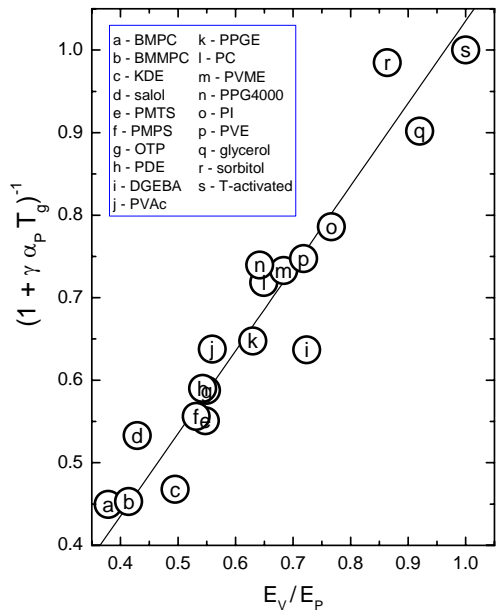

**Figure 1**

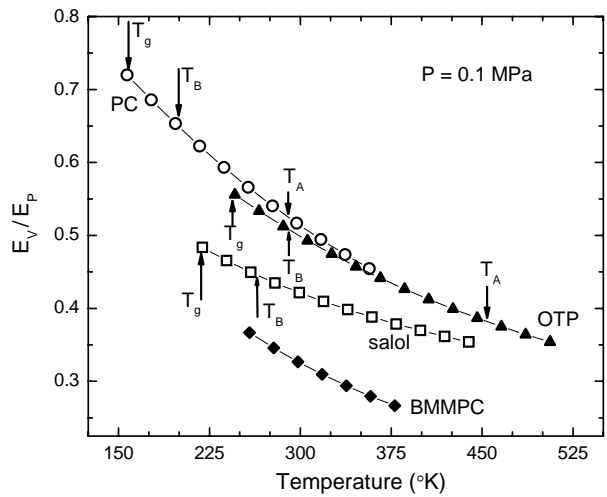

**Figure 2**

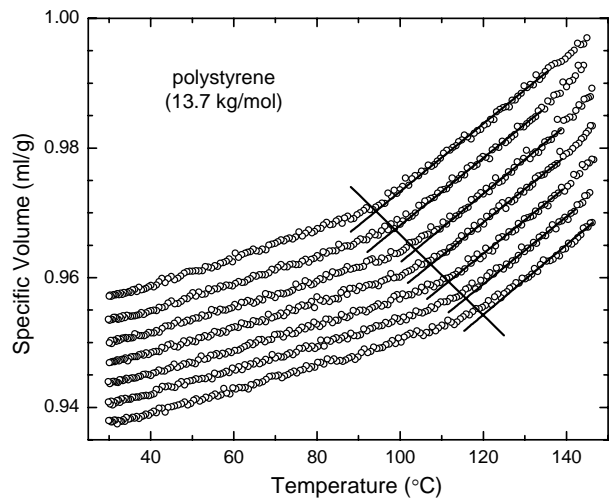

**Figure 3**

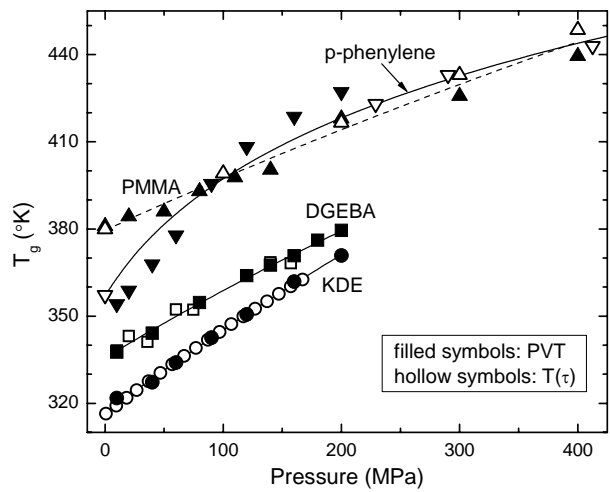

**Figure 4**

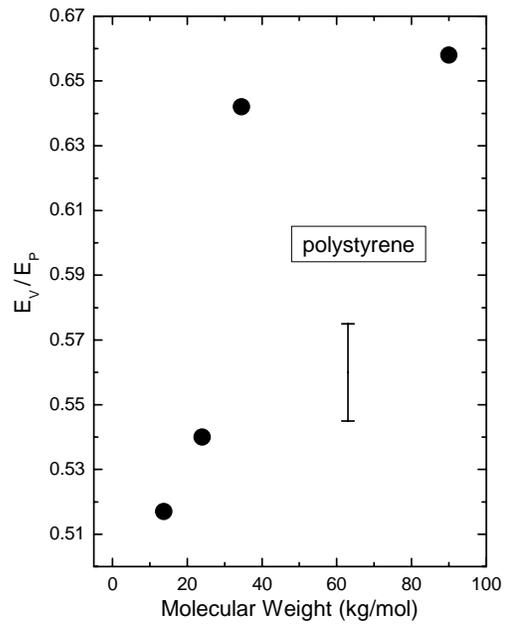

**Figure 5**